\documentclass[journal]{IEEEtran}
\usepackage{amsmath,amsfonts,amssymb}
\usepackage{algorithmic}
\usepackage{algorithm}
\usepackage{array}
\usepackage[caption=false,font=footnotesize,labelfont=rm,textfont=rm]{subfig}
\usepackage{textcomp}
\usepackage{stfloats}
\usepackage{url}
\usepackage{graphicx}
\usepackage{cite}
\usepackage{amsmath}
\usepackage{multirow}
\usepackage{array}
\usepackage{color}
\usepackage{float}
\usepackage[colorlinks,linkcolor=blue,anchorcolor=blue,citecolor=blue]{hyperref}% use this package to realize the hyperlink of ref.	-zzy2023
\usepackage{pgfplots}
\usetikzlibrary{matrix}
\usetikzlibrary{calc}
\pgfplotsset{compat = 1.18}
\begin{document}
	
	\title{Cross-Frame OTFS Parameter Estimation Based On Chinese Remainder Theorem}
	
	\author{Zhenyu Zhang, Qianli Wang,\IEEEmembership{ Member,~IEEE}, Gang Liu, \IEEEmembership{Member,~IEEE}, FeiFei Gao, \IEEEmembership{Fellow,~IEEE}, and Pingzhi Fan, \IEEEmembership{Fellow,~IEEE}
		% <-this % stops a space
		\thanks{Z. Zhang, Q. Wang, G. Liu, and P. Fan are with the Key Laboratory of Information Coding and Transmission, Southwest Jiaotong University, Chengdu 610031, China (e-mail: zzy1997@my.swjtu.edu.cn; qianli\_wang@qq.com; gangliu@swjtu.edu.cn; pzfan@swjtu.edu.cn). 
			F. Gao is with the Department of Automation, Tsinghua University, Beijing 100084, China (e-mail: feifeigao@ieee.org).
			The work is supported by the NSFC (Natural Science Foundation of China) project under Grant No. 61971359, No. 62301455 and Sichuan Science and Technology Program (Grant No. 2023ZHCG0010, No. 2023YFH0012, and No. 2023YFG0312). (Corresponding author: Qianli Wang.)	}% <-this % stops a space
	}

		\markboth{\parbox{\textwidth}{This work has been submitted to the IEEE for possible publication. \\ Copyright may be transferred without notice, after which this version may no longer be accessible.}}%
		{Shell \MakeLowercase{\textit{et al.}}: A Sample Article Using IEEEtran.cls for IEEE Journals}
	
	\IEEEpubid{}
	
	\maketitle
	
	\begin{abstract}
		Orthogonal time-frequency space (OTFS) is a potential waveform for integrated sensing and communications (ISAC) systems because it can manage communication and sensing metrics in one unified domain, and has better performance in high mobility scenarios. In practice, a target might come from far distance or with ultra-high speed. However, the max unambiguous range and max tolerable velocity of OTFS-ISAC system is limited by the unambiguous round-trip delay and Doppler shift, which are related to OTFS frame, i.e., time slots and subcarrier spacing, respectively.  
		To enlarge the sensing range, a novel OTFS cross-frame ranging and velocity estimation model as well as its corresponding method based on the Chinese remainder theorem (CRT) are proposed in this paper. 
		By designing co-prime numbers of subcarriers and time slots in different subframes, the difference in the responses of the subframes for a target can be used to estimate the distance and velocity of an out-of-range target.
		Several frame structures are further designed for specific sensing scenarios, such as target with ultra-high speed or at far distance.
		Simulation results show that the proposed method can achieve significantly better performance in NMSE compared with the classic sensing methods under the condition of same time and frequency resources.
	\end{abstract}
	
	
	\section{Introduction}
	\IEEEPARstart{I}{n} the face of increasingly congested radio resources, to develop new functionalities and enable different application scenarios like vehicle-to-everything (V2X) \cite{JPROC_V2X_2019} \cite{JPROC_V2X_2022}, unmanned aerial vehicles (UAVs) \cite{COMST_UAV_2022} \cite{MWC_UAV_2024}, smart industries \cite{rao_impact_2018}, etc., the upcoming sixth-generation (6G) of wireless communication is anticipated to possess sensing ability along with evolving communication capabilities, i.e., Integrated Sensing and Communications (ISAC) \cite{yuan_otfs_2023} \cite{lu_integrated_2024}. This promising technology enables both communication and sensing purposes to be fulfilled in a unified signal by the same transmitter and receiver in the sophisticated architecture\cite{noauthor_framework_2023}, which has also been proven to have mutual benefits on both functionalities \cite{xiong_fundamental_2023}.
	
	Waveform design plays a vital role in ISAC research.
	Some initial works, such as \cite{sturm_waveform_2011,Chiriyath_TCCN_2017,Striano_RWS_2019,Uysal_TVT_2020} take orthogonal frequency division multiplexing (OFDM) as a waveform for ISAC. 
	However, the OFDM waveform in current 5G and 5G-A systems suffers from severe inter-carrier interference (ICI) in high-mobility channels, like high-speed railways \cite{JSAC_HSR_2020,HSR_MCOM_2022} or low-earth-orbit satellites \cite{su_broadband_2019,MCOM_LEO_2021}, leading to performance degradation.
	
	The recently proposed orthogonal time-frequency space (OTFS) modulation \cite{hadani_2_2017,hadani_orthogonal_2017}, which can handle doubly-selective channels well, has been considered as one of the candidate waveforms for future wireless communication systems \cite{MCOMSTD_2024}. 
	Compared with OFDM, OTFS shows advantages in resisting the Doppler shifts, realizing smaller peak-to-average power ratio (PAPR), and reducing cyclic prefix (CP) overhead \cite{yuan_otfs_2023}. 
	Furthermore, the delay Doppler (DD) domain channel reflects the physical attributes of the propagation path and the DD domain taps correspond to the scatterers' delay and Doppler shifts \cite{TWC_channel_2005,ICCW_2019}, which leads to a common domain for communication symbols' and sensing parameters' representation \cite{wang_multi-symbol_2023}.
	Therefore, there have been many works combining OTFS with ISAC systems recently and these works can be divided into three groups, i.e., waveform design, performance analysis, and  ISAC framework.
	
	In terms of waveform design, a DFT (Discrete Fourier Transform)-Spread OTFS with superimposed pilots was elaborated for Terahertz (THz) ISAC scenarios aimed to improve the robustness of Doppler effects and reduce PAPR \cite{wu_dft-spread_2023}. Corresponding two-phase sensing parameter estimation for multiple targets as well as symbol detection algorithms with a conjugate gradient-based equalizer was also proposed for DFT-Spread OTFS ISAC systems. In \cite{cui_multi-domain_2023}, a DD and TF domain ISAC scheme based on nonorthogonal resource allocation was proposed, where communication data was arranged both on the TF and DD domain and extra degrees of freedom (DoF) was naturally achieved. An iteration-based receiver was also designed to mitigate the power domain interference. In \cite{zhang_dual-functional_2024}, the problem of ISAC waveform design was formulated as a problem of minimizing the interference terms caused by the ISAC waveform and weighted integrated sidelobe level. 
	The transmit sequence was optimized via OTFS signaling and the performance was improved by the increased DoF in local ambiguity function shaping.
	
	There have been many researches focused on the ISAC system performance analysis, including the tradeoff between communication and sensing performance in ISAC systems\cite{xiong_fundamental_2023,an_fundamental_2023,liu_deterministic-random_2023}. In the aspect of performance analysis in the OTFS-ISAC system, most works contribute to the comparison between OTFS and OFDM. An OTFS modulation-based radar system was introduced in \cite{raviteja_orthogonal_2019} and the sensing performance of OFDM and OTFS was also compared. In \cite{TWC_Gaudio_2020}, Gaudio et al. analyzed the effectiveness of OTFS modulation in the ISAC scheme and derived an approximated Maximum Likelihood algorithm for Doppler and delay estimation. Numeral results showed that OTFS have the same sensing performance as radar waveforms for dual-functional sensing and communication systems. Further, in \cite{ICC_Zhang_2023}, Zhang et al. proved that demodulation of OTFS with rectangular pulse shaping is exactly the range-Doppler matrix computing process in radar sensing, showing the connections between OTFS communication and radar sensing. 
	
	In terms of OTFS-ISAC framework. Yuan et al. investigated OTFS-based ISAC framework in vehicular networks\cite{JSTSP_Yuan_2021}, where radar sensing was used to obtain the channel state information (CSI) and assisted the downlink communication, thereby reducing the channel estimation overhead. 
	In \cite{li_novel_2022}, a novel ISAC framework was proposed based on spatial spread OTFS where the angular domain is discretized, and the author also designed a precoding scheme for communication based on the results from radar sensing. The conclusion drawn was that the power allocation should be designed to lean towards radar sensing in practical scenarios.
	In \cite{wang_multi-symbol_2023}, a compressive sensing framework for an OTFS-based ISAC system was proposed where the sensing resolution was adjustable and can be compatible with fractional channel parameters. Then an orthogonal matching pursuit (OMP)-based solver was developed and showed better performance than the matched filter method. 
	A Doppler spectrum matching assisted active sensing framework consisting of three modules was also proposed to improve the parameter estimating accuracy\cite{xia_achieving_2024}. 
	In \cite{jafri_sparse_2024}, Jafri et al. proposed an OTFS-ISAC framework in the mmWave band with a hybrid beamforming architecture. Leveraging the inherent sparsity of the signal scattering environment, a two-stage Bayesian learning (BL)-based procedure was conceived for transceiver design, radar target parameter, and wireless channel estimation which results in improved estimation accuracy.
	In \cite{yang_barycenter_2024,liu_two-step_2024,liu_amplitude_2024}, an amplitude barycenter calibration algorithm based delay-Doppler signaling frameworks were proposed for OTFS-ISAC system, where multiple pilots were used and receive window design was also considered. Then the sensing performance was improved without broadening the bandwidth.
	In \cite{Zegrar_TCOMM_2024}, authors proposed a high delay and Doppler resolution framework for OTFS-ISAC systems, where a single OTFS carrier was used as a pilot to estimate fractional channel parameters and the constraints in time and frequency bandwidth were relaxed. The simulation and real experimental results showed that precise sensing information can be obtained with the proposed framework. 
	In \cite{ouyang_rfsoc-based_WCNC}, a scalable OTFS-ISAC prototyping platform based on RFSoC (Radio Frequency System on Chip) was introduced. The platform featured an efficient frame structure with low-complexity channel estimation and equalization algorithms for communication. Additionally, the synchronization preamble and the peak pilot were utilized for sensing.
	
	Although these works have achieved good performance, they all assumed that the scatters in the scenarios are in a certain limited region, which may not hold in practice. The region is related to the OTFS frame size, i.e., limited distance and velocity owing to time duration and subcarrier spacing, respectively.  
	This is called the crystallization condition in \cite{Mohammed_MBITS_2022}, and when it is not satisfied, the channel will be unpredictable. The distance and velocity of the target will also be unpredictable. This may happen in scenarios in which the targets are with ultra-high speed or at far distance.
	This phenomenon is not prominent in traditional communication or radar systems. For communication systems, the transmission range is certain, and the CP is usually used in OFDM to evade the problem. While in radar systems, applications or targets are certain thus the coherent processing interval, i.e., maximum range, is usually fixed.
	However, in ISAC systems, for the compatibility of dual functionalities, the communication parameters time slot $T$ and subcarrier spacing $\Delta f$ as well as the signal frame structure should not be changed frequently. But the applications and targets in the scenarios may be various, which may lead to ambiguous sensing results. 
	
	Motivated by co-prime antenna array signal processing \cite{Vaidyanathan_TSP_2011,Li_TSP_2019}, which enlarges the aperture of the array through the Chinese remainder theorem (CRT) \cite{Ding_CRT_1996}, a novel OTFS cross-frame ranging and velocity estimation model as well as its corresponding method are proposed in this paper.
	Be different from works of \cite{Vaidyanathan_TSP_2011,Li_TSP_2019}, which aim to increase resolution, co-prime OTFS frames proposed in this paper are designed to acquire a DoF gain thereby breaking the unambiguous range and tolerable velocity limit.
	The main contributions of this paper can be summarized as follows:
	\begin{itemize}
		\item 
		We show the periodic effect of channel parameters and the unambiguous range and tolerable velocity limitation in OTFS-ISAC systems. The range-velocity profile of scatterers needs to be confined within a specific region, otherwise the results will be ambiguous. 
		A novel OTFS cross-frame ranging and velocity estimation model is derived.
		\item A novel framework based on the OTFS cross-frame ranging and velocity estimation model is proposed.
		In this framework, co-prime subcarrier numbers and time slots are used in different subframes. 
		Then the max unambiguous delay and Doppler taps can be significantly increased from $N_s$ and $M_s$ to near $(N_s)^{F}$ and $(M_s)^{F}$, respectively. $F$ is the number of subframes.	
		
		\item Several frame structures are designed according to specific sensing scenarios.
		Co-prime subcarriers and guard intervals are designed to handle excessive delay for target at far distance. Besides, co-prime time slots and guard intervals are designed to handle excessive Doppler for target with ultra-high speed.
		A multi-frame structure is also proposed to expand the both velocity and distance range at the expense of removing information symbols for communication.
	\end{itemize}
	
	The remainder of this paper is organized as follows. Sec. \ref{sec2} introduces the OTFS ISAC system, derives the multicarrier signal periodicity, and analyzes the range and velocity limitation in the OTFS-ISAC system. In Sec. \ref{sec3}, the cross-frame ranging and velocity estimation model as well as its method are proposed to expand the sensing range. Sec. \ref{sec4} considers frame structure for cross-frame processing in practical ISAC systems.  Sec. \ref{sec5} gives the simulation results in different conditions. Finally, Sec. \ref{sec6} concludes the paper.

	\section{OTFS ISAC Model for Out-of-range Target}\label{sec2}
	Previous works usually assume that the delay and Doppler of the scatterers are in a certain range\cite[Chapter 4]{hong_delay-doppler_book_2022}, or the sensing ability of the ISAC system is restricted in an unambiguous range \cite{raviteja_orthogonal_2019}, 
	\begin{equation}
		\label{eq15}
		\begin{aligned}
			\tau_\text{max}<T \quad \text{and} \quad \nu_\text{max}<\Delta f/2,
		\end{aligned}
	\end{equation}
	where $T$, $\Delta f$, $\tau_\text{max}$ and $\nu_\text{max}$ represent the time slot, subcarrier spacing, the maximum delay and Doppler shift in the propagation environment, respectively. 
	In Zak-OTFS, this assumption is related to crystalline condition \cite{mohammed_otfs_2023}, which can decide whether the channel is predictable. 
	
	\begin{figure}
		\centering
		\includegraphics[width = 0.48\textwidth]{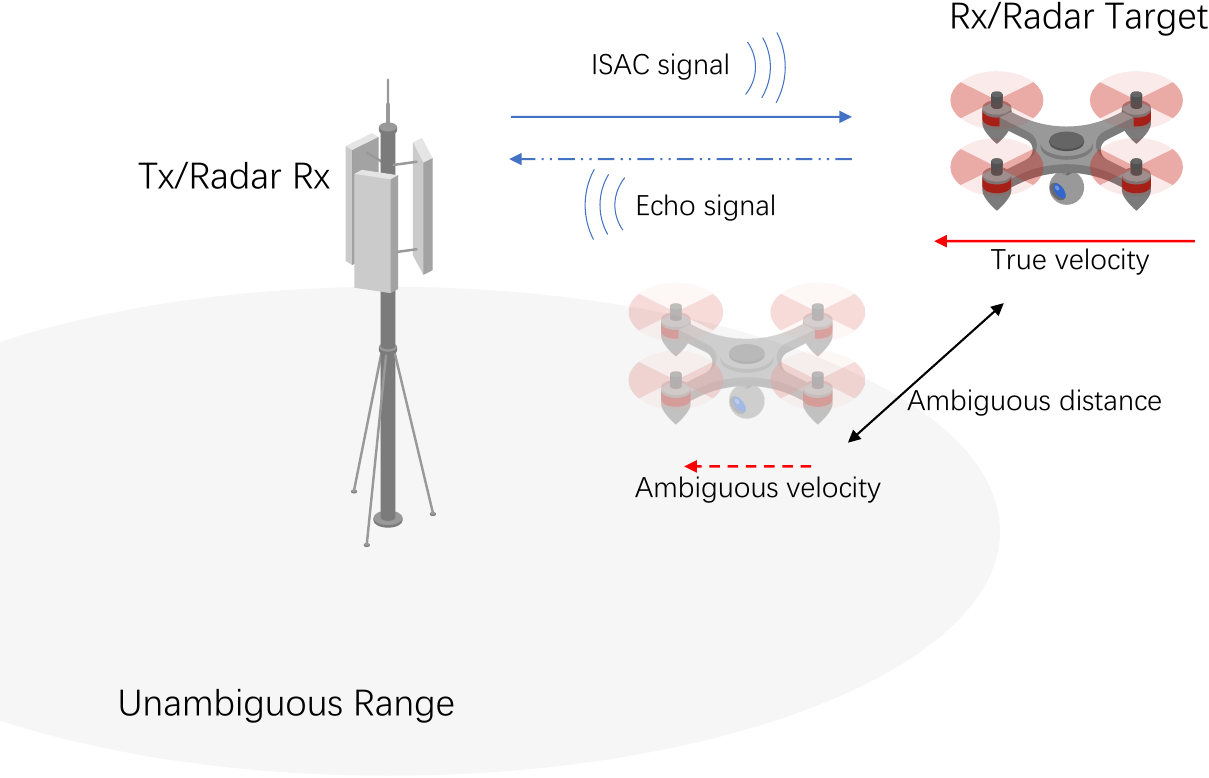}
		\caption{Range and Velocity limits due to the periodicity of OTFS.}
		\label{fig1}
	\end{figure}
	
	But in practice, a target might be at far distance or with ultra high speed, as shown in 
	Fig. \ref{fig1}.  A UAV target comes from far distance into the coverage of the station and it is with ultra high speed. 
	An OTFS-based ISAC system with a single antenna at both the transmitter and receiver is considered. 
	The ISAC channel can be regarded as a doubly-selective fading channel \cite{hlawatsch_wireless_2011},
	\begin{equation}
		\label{EqChannel}
		h(\tau,\nu)=\sum_{i=1}^Ph_i\delta(\tau-\tau_i)\delta(\nu-\nu_i),
	\end{equation}
	where $P$ is the number of scatterers, $h_i$ is the complex gain of the $i$-th path, $\tau_i$ and $\nu_i$ are the delay and Doppler shift of the $i$-th scatterer, respectively. 
	
	After the channel is convoluted with the transmitting time domain signal $ s(t)$, which is the same as that in \cite[Eq. (3)]{raviteja_interference_2018}, the received signal can be expressed as:
	\begin{equation}
		\label{EqReceivedSignal}
		r(t)= \sum_{i=1}^{P} h_i s(t-\tau_i)e^{j2\pi(t-\tau_i)\nu_i}.
	\end{equation}
	
	For radar sensing, the received signal has the same format as Eq. \eqref{EqReceivedSignal}, but $h_i$ is decided by the radar cross section (RCS) of the $i$-th scatterer. $\tau_i$ and $\nu_i$ represent the round-trip propagation shift of $i$-th scatterer for mono-static radar, i.e., $\tau_i = 2r_i/c$ and $\nu_i =2v_i f_m/c$, where $r_i$ and $v_i$ are the range and velocity of the $i$-th scatterer, $f_m$ is the modulation frequency, and $c$ is the speed of light. 
	After matched filtering, the received signal is:
	\begin{equation}
		% \label{eq4.1}
		Y(t,f)=A_{r,g_{\mathrm{rx}}}(f,t)\triangleq\int r\left(t'\right)g_{\mathrm{rx}}^{*}(t'-t)e^{-j2\pi f(t'-t)}dt',
	\end{equation}
	
	where $g_{\mathrm{rx}}(t)$ is the impulse response of the receiver filter. 
	Then received signal is sampled at the receiver with a sampling period of time slot $T$ and a subcarrier spacing $\Delta f$,
	\begin{equation}
		\label{EqSampling}
		\hat{Y}[n,m]=Y(f,t)|_{f=m\Delta f,t=nT},
	\end{equation}
	where $n = 0,1,...,N-1, m = 0,1,...,M-1$. $M, N$ is the number of subcarriers and time slots, respectively.
	
	In this paper, we only consider the case of one target with out-of-range parameters, i.e., $P = 1$.
	When multiple targets with out-of-range parameters in the channel, the situation will be quite different, as the received signals will superimpose and spread over the entire TF domain, making it difficult to estimate. It will be discussed in our future work.
	The target's out-of-range delay  $\tau_i$ and Doppler $\nu_i$ can be written as an out-of-range part and an in-range part:
	\begin{equation}
		% \label{eq4.3}
		\begin{aligned}
			&\tau_i=\alpha_i T+\hat{\tau}_{i},\quad\alpha_i \in \mathbb{Z},\quad\hat{\tau}_{i}<T,\\
			&\nu_i=\beta_i \Delta f+\hat{\nu}_{i},\quad\beta_i \in \mathbb{Z},\quad\hat{\nu}_{i}<\Delta f.
		\end{aligned}
	\end{equation}
	In this case, the sampling range of \eqref{EqSampling} would be adjusted for collecting all the received signal.
	Therefore the DD domain received signal can be expressed as:
	\begin{equation}
		\label{EqDDSignal}
		\begin{aligned}
			&y[k,l]=\frac{1}{\sqrt{NM}}h_{i}e^{{-j2\pi\hat{\tau}_{i}(\beta_{i}\Delta f+\hat{\nu}_{i})}}\sum_{n=0}^{N-1}\sum_{n'=0}^{M-1}\sum_{m=0}^{N-1}\sum_{m'=0}^{M-1}\\
			&e^{{j2\pi\hat{\tau}_{i}\cdot nT}}X[n^{\prime},m^{\prime}]e^{{j2\pi m^{\prime}\Delta f(nT-n^{\prime}T-\hat{\tau}_{i})}}\\
			&A_{{g_{{\mathrm{tx}}},g_{{\mathrm{rx}}}}}((n-n^{\prime})T-\hat{\tau}_{i},(m-m^{\prime})\Delta f-\hat{\nu}_{i})e^{{-j2\pi(\frac{nk}{N}-\frac{ml}{M})}},
		\end{aligned}
	\end{equation}
	
	The complete derivation is given in the Appendix. As shown in Eq. (\ref{EqDDSignal}), the out-of-range delay $\alpha_i T$ shows periodicity and has no effect on the received signal.
	While the out-of-range Doppler $\beta_i \Delta f$ only affects the phase of channel coefficients, which is combined in the $h_i$.
	
	If ideal pulse is assumed, from \cite{raviteja_interference_2018}, the received DD domain signal can be expressed as:
	\begin{equation}
		\label{eq9}
		\begin{aligned}
			y[k,l]=&\tilde{h}_i \sum_{k'=0}^{N-1}\sum_{l'=0}^{M-1}\\
			&w_\nu(N,k-k'-k_{\nu_i}) w_\tau(M,l-l'-l_{\tau_i}).
		\end{aligned}
	\end{equation}
	where  $x[k,l],k=0,\dots,N-1, l=0,...,M-1$ is DD domain transmit signal.
	$\tilde{h}_i=h_{i} e^{-j2\pi\frac{k_{\nu_i}\hat{l}_{\tau_i}}{NM}}$. $k_{\nu_i}$ and $l_{\tau_i}$ are the normalized delay and Doppler taps for the $i$-th scatterer,
	\begin{equation}
		\label{eq10}
		\begin{aligned}
			k_{\nu_i}=&\nu_i NT = \beta_i N+ \hat{\nu_i}T,
			\\
			l_{\tau_i}=&\tau_i M\Delta f = \alpha_i M+ \hat{\tau_i}\Delta f.
		\end{aligned}
	\end{equation}
	The in-range part of delay and Doppler taps are $\hat{k}_{\nu_i}= \hat{\nu_i}T$ and $\hat{l}_{\tau_i}=\hat{\tau_i}\Delta f$, respectively.
	$w_\nu$ and $w_\tau$ are the sampling functions, 
	\begin{equation}
		\label{eq11}
		\begin{aligned}
			w_\nu(N,k-k'-k_{\nu_i})=&\frac{1}{N}e^{-j\pi (k-k'-k_{\nu_i})\frac{N-1}{N}}\\
			&\cdot \frac{\sin(\pi (k-k'-k_{\nu_i}))}{\sin(\pi \frac{(k-k'-k_{\nu_i})}{N})},\\
			w_\tau(M,l-l'-l_{\tau_i})=&\frac{1}{M}e^{j\pi (l-l'-l_{\tau_i})\frac{M-1}{M}}\\
			&\cdot \frac{\sin(\pi (l-l'-l_{\tau_i}))}{\sin(\pi \frac{(l-l'-l_{\tau_i})}{M})}.
		\end{aligned}
	\end{equation}
	
	Though out-of-range delay and Doppler are assumed, it can be found that the input-output relationship in the DD domain is in coincidence with that of \cite[Eq. (14)]{raviteja_interference_2018} because of the periodicity.
	
	However, the periodicity leads to an ambiguity problem in representing out-of-range delay and Doppler. 
	$w_\nu$ and $w_\tau$ are periodic functions with periods $N$ and $M$, respectively. That is, $\forall a, b\in \mathbb{Z}$, we have:
	\begin{equation}
		\label{eq12}
		\begin{aligned}
			w_\nu(N,k-k'-k_{\nu_i}-a\cdot N) &= w_\nu(N,k-k'-k_{\nu_i}),\\
			w_\tau(M,l-l'-l_{\tau_i}-b\cdot M)&= w_\tau(M,l-l'-l_{\tau_i}),
		\end{aligned}
	\end{equation} 
	
	Then the received signal Eq. (\ref{eq9}) is equivalent to:
	\begin{equation}
		\label{eq13}
		\begin{aligned}
			y[k,l]=&\sum_{k'=0}^{N-1}\sum_{l'=0}^{M-1}x[k',l']\cdot
			\\&\tilde{h}_i\cdot w_\nu(N,k-k'-k_{\nu_i}-a\cdot N) \\&\cdot w_\tau(M,l-l'-l_{\tau_i}-b\cdot M),
		\end{aligned}
	\end{equation} 
	Thus the true distance and velocity of the target will be unpredictable.
	Without considering noise, the estimated delay and Doppler taps $\tilde{k}_i$ and $\tilde{l}_i$ by Eq. \eqref{eq13} can be expressed as:
	\begin{subequations}
		\label{eq14}
		\begin{align}
			\tilde{k}_i &= k_{\nu_i}-a\cdot N,\label{eq14.1}\\
			\tilde{l}_i &= l_{\tau_i}-b\cdot M.\label{eq14.2}
		\end{align}
	\end{subequations}
	where $k_{\nu_i}$ and $l_{\tau_i}$ are the true delay and Doppler taps corresponding to the targets. 
	
	It could be found that the classic OTFS-ISAC system \cite{raviteja_orthogonal_2019} will estimate an ambiguous distance and velocity of the target due to the periodicity of OTFS, as shown in Fig. \ref{fig1}.
	This issue is not prominent in traditional communication or radar systems. As in communication systems, the transmission range is usually certain and CP is used. While in radar systems, applications or targets are certain thus the max unambiguous range and tolerable velocity is usually fixed. 
	However, in ISAC systems, the communication parameters $T$ and $\Delta f$ as well as the signal frame structure should not be changed frequently. But the applications and targets may be various, which may lead to ambiguous sensing results. 
	Therefore, a novel model as well as the estimation method are proposed to expand the sensing range.

	\section{Cross-Frame Parameter Estimation based on Chinese Remainder Theorem}\label{sec3}
	To extend the unambiguous delay and Doppler range, a cross-frame parameter estimation method is proposed in this section. 
	We will first assume that each OTFS frame is with a single pilot for pure sensing scenes. Its extension for the ISAC system will be discussed in the next section.
	
	\subsection{Cross-Frame Formulation with Two Frames}
	
	For the physical attributes represented in the DD channel are slowly changed \cite{TWC_channel_2005}, we can assume a multipath channel with its parameters $h_i$, $\nu_i$ and $\tau_i$ being constant throughout two OTFS frames. 
	To obtain a larger sensing range, we use two or more consecutive subframes with different subcarrier spacing and time slot lengths.
	By processing these subframes independently, delays and Dopplers with ambiguity can be obtained. Then we will combine these results for solving ambiguity.
	
\begin{figure}[H]
		\centering
		\includegraphics[width = 0.48\textwidth]{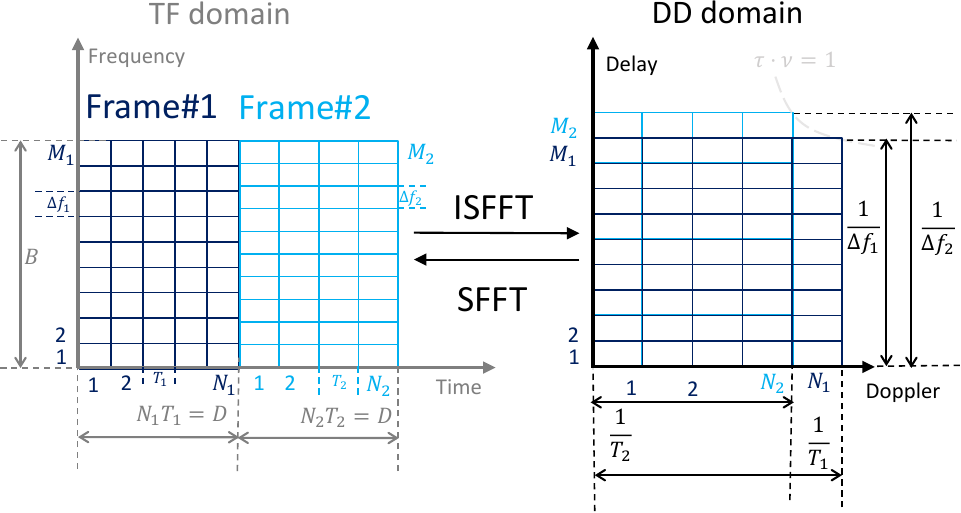}
		\caption{Consecutive OTFS frames.}
		\label{fig2}
	\end{figure}

	Assume that the subframes have the same duration and bandwidth, i.e., $D$ and $B$,  respectively. The numbers of subcarriers and time slots of the subframes are $M_1,N_1$ for Frame 1 and $M_2,N_2$ for Frame 2, respectively.
	Thus the TF domain and DD domain of the subframes are shown in Fig. \ref{fig2}.
	The subcarrier spacing and time slot length in the first and second frames are $\Delta f_1$, $T_1$ and $\Delta f_2$, $T_2$, respectively.
	
	Note that, $N_1$ and $N_2$, $M_1$ and $M_2$ should be mutually prime, i.e., $\mathrm{GCD}(M_1, M_2)=1$, $\mathrm{GCD}(N_1, N_2)=1$, where $\mathrm{GCD}$ represents the greatest common divisor.
	
	Denote the DD domain transmit and received signals of frame 1 and frame 2 as $x_1[k_1,l_1], y_1[k_1,l_1],k_1=0,\dots,N_1-1, l_1=0,...,M_1-1$ and $x_2[k_2,l_2],y_2[k_2,l_2],k_2=0,\dots,N_2-1, l_2=0,...,M_2-1$, respectively. 
	If there is only one single pilot in each frame, i.e., $x_1[k_{p_1},l_{p_1}]\neq 0$, $x_2[k_{p_2},l_{p_2}]\neq 0$, and all other symbols are set zero in the DD domain, based on analysis in Sec. \ref{sec2}, the received signal with ideal pulses can be expressed as:
	\begin{equation}
		\label{eq17}
		\begin{aligned}
			y_1[k_1,l_1]&=x_1[k_{p_1},l_{p_1}]\cdot \\
			&\tilde{h}_i w_\nu(N_1,k-k_{p_1}-k_{\nu_i}) w_\tau(M_1,l-l_{p_1}-l_{\tau_i}),\\
			y_2[k_2,l_2]&=x_2[k_{p_2},l_{p_2}]\cdot \\
			&\tilde{h}_i w_\nu(N_2,k-k_{p_2}-k_{\nu_i}) w_\tau(M_2,l-l_{p_2}-l_{\tau_i}),
		\end{aligned}
	\end{equation}
	where $x_1[k_{p_1},l_{p_1}]$ and $x_2[k_{p_2},l_{p_2}]$ represent the pilot in the Frame 1 and in the Frame 2, respectively. $w_\nu$ and $w_\tau$ are the sampling functions given in Eq. (\ref{eq11}). 
	
	If the delay and Doppler taps are in the range as that of Eq. \eqref{eq15}, then the responses of the two subframes are the same. The distance and velocity can be extracted directly.
	When there are out-of-range delay and Doppler taps, it can be found from Eq. \eqref{eq17} that the responses of the two subframes are different because of their different subcarrier spacings and time slot lengths. 
	In the following subsection, we will elaborate on the cross-frame excessive parameter estimation method to extract the true distance and velocity of a target. 
	
	\subsection{Extend Sensing Limit with CRT}\label{sec3.1}
	For simplicity, assume that the channel is with integer delay and Doppler taps. 
	The fractional parameters will not affect the use of our method because the fractional parts always meet Eq. \eqref{eq15}. We can just solve the ambiguity for the integer part. Then methods, like that in \cite{wang_multi-symbol_2023}\cite{Zegrar_TCOMM_2024} can be used to solve fractional parameters.
	
	When the channel parameters fail to meet Eq. (\ref{eq15}), then the normalized delay and Doppler taps in DD domain will satisfy the following conditions:
	\begin{equation}
		\label{eq18}
		\begin{aligned}
			&l_{\tau_i} > \max\{M_1,M_2\},\quad l_{\tau_i}\in \mathbb{Z} ,\\
			&|k_{\nu_i}| > \max\{N_1,N_2\}/2,\quad k_{\nu_i}\in \mathbb{Z}.
		\end{aligned}
	\end{equation}
	Note that the values of $l_{\tau_i}$ and $k_{\nu_i}$ are irrelevant to the parameter settings of subframes (Eq. \eqref{eq10}) because the subframes have same durations $NT$ and bandwidths $M\Delta f$.
	
	Then, Eq (\ref{eq17}) can be expressed as:
	\begin{equation}
		\label{eq19}
		\begin{aligned}
			y_1[k_1,l_1]=&x_1[k_{p_1},l_{p_1}]\tilde{h}_i\cdot \\
			&\delta[[k_1-k_{p_1}-k_{\nu}]_{N_1}] \delta[[l_1-l_{p_1}-l_{\tau}]_{M_1}],\\
			y_2[k_2,l_2]=&x_2[k_{p_2},l_{p_2}]\tilde{h}_i\cdot \\
			&\delta[[k_2-k_{p_2}-k_{\nu}]_{N_2}] \delta[[l_2-l_{p_2}-l_{\tau}]_{M_2}],\\
		\end{aligned}
	\end{equation}
	where $\delta[\cdot]$ and $[\cdot]_N$ represent the Kronecker delta function and modulo $N$ operation, respectively. 
	$l_{\tau_i}$ and $k_{\nu_i}$ are the true normalized delay and Doppler tap corresponding to the scatterer.
	
	We can intuitively estimate the delay and Doppler shift of the scatterer in each frame as that in \cite{raviteja_embedded_2019}, denoted as $\hat{l}_{\tau_1}$, $\hat{k}_{\nu_1}$ and $\hat{l}_{\tau_2}$, $\hat{k}_{\nu_2}$ for the two subframes, respectively. 

	When the parameters are out-of-range, then $\hat{l}_{\tau_1}\neq\hat{l}_{\tau_2}$ and they are both smaller than $l_{\tau_i}$. Similarly, $\hat{k}_{\nu_1}\neq\hat{k}_{\nu_2}$ and they are smaller than $k_{\nu_i}$.
	As $N_1$, $N_2$ and $M_1$, $M_2$ are co-prime, due to the periodicity of OTFS, we have the following linear congruence equations:
	\begin{subequations}
		\label{eq21}
		\begin{align}
			\tilde{k}_{\nu_i}  &\equiv \hat{k}_{\nu_1} \pmod{N_1},\label{eq3.61}\\
			\tilde{k}_{\nu_i}  &\equiv \hat{k}_{\nu_2} \pmod{N_2},\label{eq3.62}\\
			\tilde{l}_{\tau_i}  &\equiv \hat{l}_{\tau_1} \pmod{M_1},\label{eq3.63}\\
			\tilde{l}_{\tau_i}  &\equiv \hat{l}_{\tau_2} \pmod{M_2},\label{eq3.64}
		\end{align}
	\end{subequations}
	where $\tilde{k}_{\nu_i}$ and $\tilde{l}_{\tau_i}$ are the estimated parameters corresponding to the true delay and Doppler. $a \equiv b \pmod{c}$ means that $(a \mod c) = (b \mod c)$. 
	The linear congruence equations can be solved by the Chinese Remainder Theorem (CRT) \cite{Ding_CRT_1996}, written as:
	\begin{subequations}
		\begin{align}
			&\tilde{k}_{\nu_i}= (\hat{k}_{\nu_1} N_1 [N_2^{(-1)}]_{N_1} + \tilde{k}_{\nu_2} N_2 [N_1^{(-1)}]_{N_2}), \label{eq18a}\\
			&\tilde{l}_{\tau_i}= (\hat{l}_{\tau_1} M_1 [M_2^{(-1)}]_{M_1} + \hat{l}_{\tau_2} M_2  [M_1^{(-1)}]_{M_2}), \label{eq18b}
		\end{align}
	\end{subequations}
	where the notation $[x^{(-1)}]_y$ refers to the multiplicative inverse of $x$ modulo $y$ when $x$ and $y$ are relatively prime. They can be calculated through Extended Euclidean Algorithm\cite{cormen_introduction_2001}. Then the max unambiguous delay and Doppler taps become  
	\begin{equation}
		\label{eq23}
		\begin{aligned}
			\tilde{l}_{\tau_i} &\in [0,M_1M_2-1],\\ %\quad \hat{l}_{\tau}\in \mathbb{Z}
			|\tilde{k}_{\nu_i}| &\in [0, (N_1N_2-1)/2].
		\end{aligned}
	\end{equation}
	
	\subsection{Extend Sensing limit with Multiple Frames}\label{sec3.3}
	From the above analysis in Sec. \ref{sec3.1}, the two co-prime frames can extend the max unambiguous range and tolerable velocity of the OTFS system. Intuitively, we can further expand the sensing limit by increasing the number of frames with co-prime time slots or subcarriers.
	
	\begin{figure}[t]
		\centering
		\includegraphics[width = 0.45\textwidth]{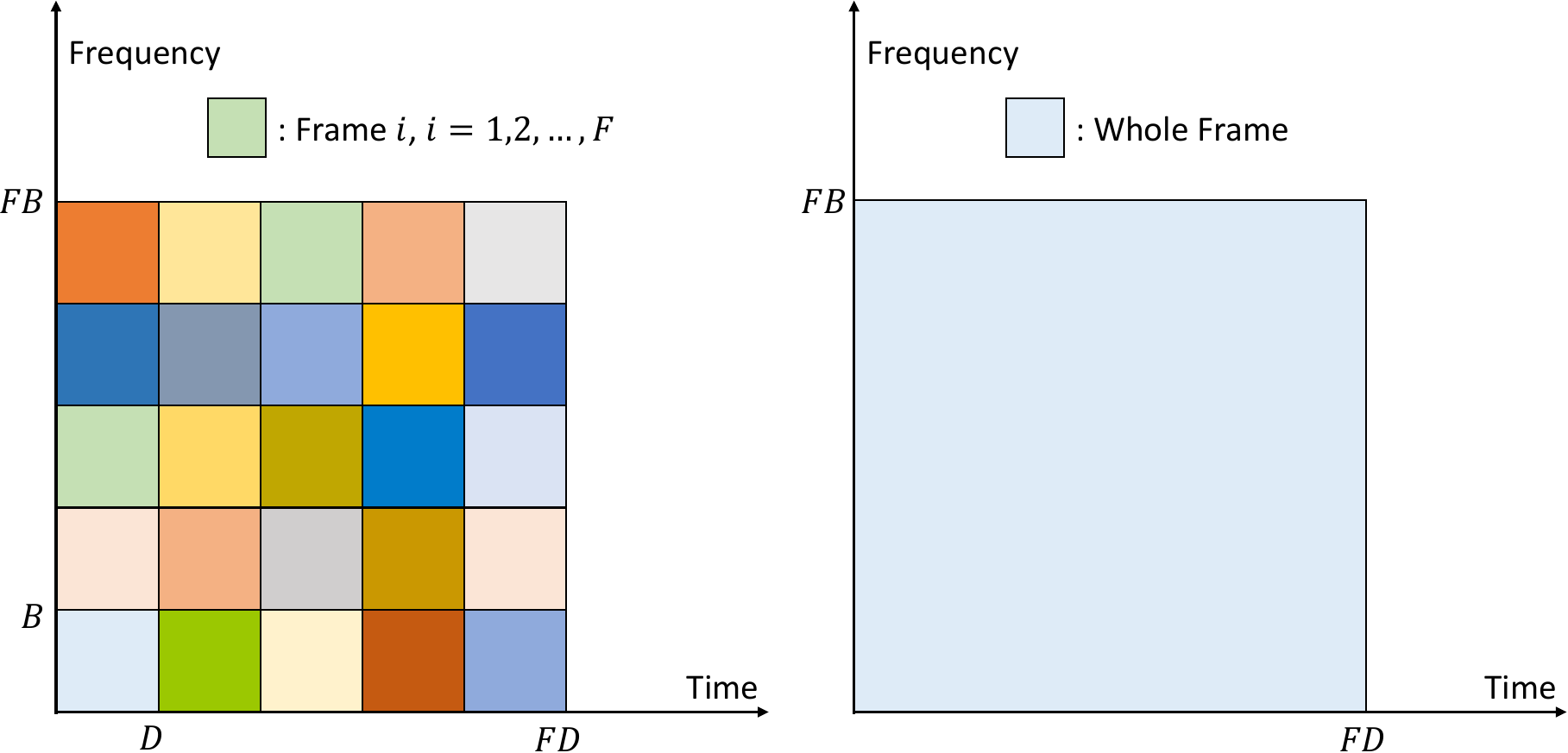}
		\caption{Multiple frames OTFS.}
		\label{fig3}
	\end{figure}
	
\begin{figure}[b]
		\centering
		\includegraphics[width = 0.48\textwidth]{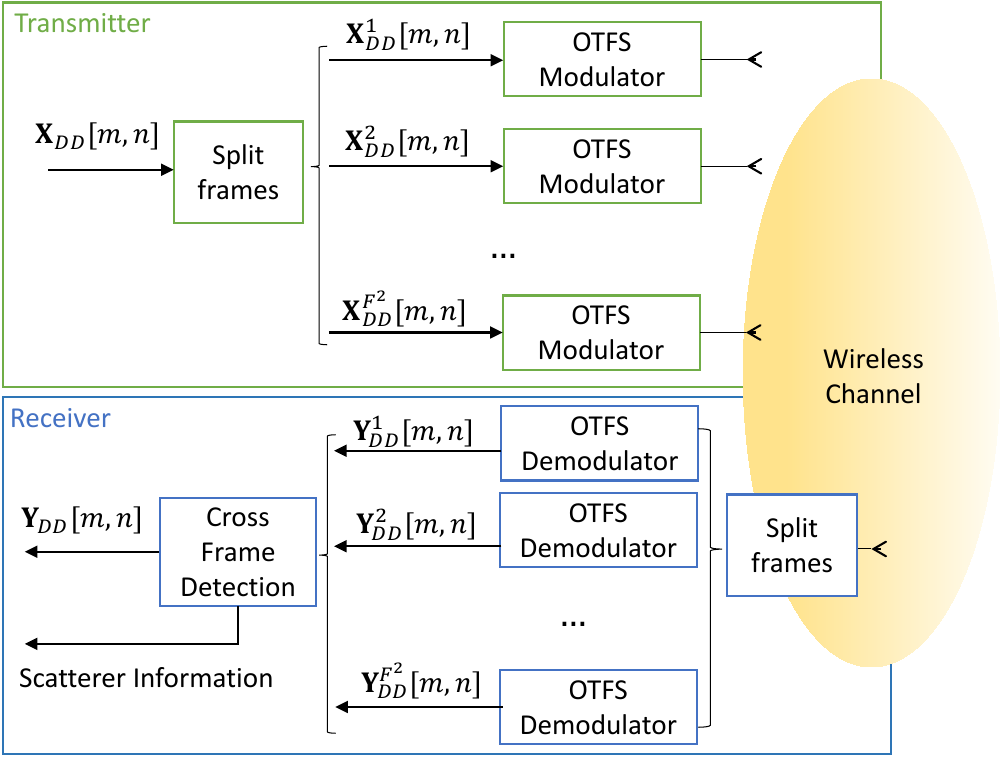}
		\caption{Cross-Frame OTFS System.}
		\label{fig4}
	\end{figure}

	Assume that a whole OTFS frame is divided into $F^2$ subframes, where time and frequency are equally divided into $F$ parts so that every subframe has the same time-frequency bandwidth. In each subframe, only one pilot is used without data symbols, and the channel remains unchanged in all subframes. The frame structure is shown in Fig. \ref{fig3} and the corresponding multi-frame OTFS system is shown in Fig. \ref{fig4}, which can represent both mono- and bi-static sensing.
	
	Denote the subcarriers and time slots in the $i$-th subframe by $N_i$ and $M_i$, $i=1,2,\dots, F$, we assume that the number of subcarriers and time slots in each subframe to be co-prime, i.e., $\mathrm{GCD}(N_i,N_j)=1$, $\mathrm{GCD}(M_i,M_j)=1$, $i\neq j$.
	After matched filter, by analyzing the received signal in each subframe, we can write the linear congruence equations of multiple subframes like Eq. (\ref{eq21}):
	\begin{equation}
		\label{eq24}
		\begin{aligned}
			\tilde{k}_{\nu_i}  &\equiv \hat{k}_{\nu_1} \pmod{N_1},\\
			\tilde{k}_{\nu_i}  &\equiv \hat{k}_{\nu_2} \pmod{N_2},\\
			& \dots\\
			\tilde{k}_{\nu_i}  &\equiv \hat{k}_{\nu_{F}} \pmod{N_{F}},\\
		\end{aligned}
	\end{equation}
	and 
	\begin{equation}
		\label{eq25}
		\begin{aligned}
			\tilde{l}_{\tau_i}  &\equiv \hat{l}_{\tau_1} \pmod{M_1},\\
			\tilde{l}_{\tau_i}  &\equiv \hat{l}_{\tau_2} \pmod{M_2},\\
			& \dots\\
			\tilde{l}_{\tau_i}  &\equiv \hat{l}_{\tau_{F}} \pmod{M_{F}}.\\
		\end{aligned}
	\end{equation}
	where $\hat{k}_{\nu_i}$ and $\hat{l}_{\tau_i}$ are the estimated delay and Doppler taps from the $i$-th subframe. Then the actual delay and Doppler taps $\tilde{k}_{\nu_i}$ and $\tilde{l}_{\tau_i}$ can be estimated by:
	\begin{equation}
		\label{eq26}
		\begin{aligned}
			\tilde{k}_{\nu_i} &= \sum_{i=1}^{F} \hat{k}_{\nu_i} n_i [n_i^{(-1)}]_{N_1},\\
			n_i &= \prod_{j=1, j\neq i}^{F} N_j,\\
		\end{aligned}
	\end{equation}
	and 
	\begin{equation}
		\label{eq27}
		\begin{aligned}
			\tilde{l}_{\tau_i} &= \sum_{i=1}^{F} \hat{l}_{\tau_i} m_i [m_i^{(-1)}]_{M_i},\\
			m_i &= \prod_{j=1, j\neq i}^{F} M_j.
		\end{aligned}
	\end{equation}
	The corresponding unambiguous range of delay and Doppler can be extended to:
	\begin{equation}
		\label{eq28}
		\begin{aligned}
			\tilde{l}_{\tau_i} &\in [0,\mathrm{LCM}(M_1,\dots,M_{F})-1],\\
			|\tilde{k}_{\nu_i}| &\in [0, (\mathrm{LCM}(N_1,\dots,N_{F})-1)/2],
		\end{aligned}
	\end{equation}
	where $\mathrm{LCM}$ represents the least common multiple.
	
	Note that, the significant enlargement of the unambiguous delay and Doppler of the proposed multiple frames is at the cost of performance desegregation of resolution compared with one whole frame with the same time-frequency consumption. But if we just want a coarse estimation of a target with far distance or ultra-high speed, this linear cost in time-bandwidth width product ($BD$) returns exponential rewards in the unambiguous delay and Doppler taps.
	The unambiguous delay, Doppler, max unambiguous range, tolerable velocity, and parameter estimation resolution are summarized in Table \ref{tab1}, where we assume that $ N_1,\dots, N_{F} $ and $ M_1,\dots, M_{F} $ are close enough so that they can be considered approximately equal, i,e., $N_1 \approx N_2 \approx \ldots \approx N_{F} \approx N_s$ and $M_1 \approx M_2 \approx \ldots \approx M_{F} \approx M_s$. $F$ is the subframes numbers which can be chosen according to the applications.
	
	\begin{table}
		\caption{Analysis of Two Frameworks}
		\label{tab1}
		\centering
		\begin{tabular}{|c|c|c|}
			\hline
			& Multiple frame & Whole frame \\
			&  (Proposed framework) & (Original OTFS)\\
			\hline
			Delay resolution & $1/B$ & $1/(B\cdot F)$ \\
			\hline
			Doppler resolution & $1/D$ & $1/(D \cdot F)$ \\
			\hline
			Unambiguous delay & $(M_s)^{F}/B$ & $M_s/B$ \\
			\hline
			Unambiguous Doppler & $(N_s)^{F}/D$ & $N_s/D$ \\
			\hline
			Max unambiguous range & $c(M_s)^{F}/B$ & $cM_s/B$ \\
			\hline
			Max tolerable velocity & $c(N_s)^{F}/(Df_c)$ & $cN_s/(Df_c)$ \\
			\hline
		\end{tabular}
	\end{table}
	
	Table \ref{tab1} shows that the proposed cross-frame processing framework can enhance the unambiguous delay and Doppler taps from $N_s$ and $M_s$ to nearly $(N_s)^{F}$ and $(M_s)^{F}$, respectively.
	Therefore the max unambiguous range and max tolerable velocity can be extended to $\frac{c(M_s)^{F}}{B}$ and $\frac{c(N_s)^{F}}{Df_c}$, respectively, where $c$ is the speed of light and $f_c$ is the carrier frequency.
	Note that, to keep low complexity and latency, the subframes are processed independently to obtain in-range delay and Doppler taps, respectively. Thus the resolution of delay and Doppler are $1/B$ and $1/D$, respectively, which are worse compared with that of using whole frame. 

	\section{Frame Structure Design For ISAC in Specific Scenarios}\label{sec4}
	Although the frame structure proposed in Sec. \ref{sec3} can extend sensing range, it is assumed that all frames carry no data symbols.
	If we want to transmit data and detect targets simultaneously, the symbols used for communication should not interfere with the response of sensing. 
	This condition cannot be satisfied if delay and Doppler are both out-of-range, in which case the response of sensing in the interested frame may be at any place, interfering with the communication symbols.
	Therefore, we can only extend the range of delay or Doppler when transmitting data simultaneously and the frame structure should be designed. 
	
	\subsection{Extend Delay Range in OTFS-ISAC System with CRT}
	Assume the target is at a far distance with in-range velocity, i.e., only the delay is out of range. Two OTFS frames are used to estimate out-of-range delay and in-range Doppler with the number of subcarriers and time slots as $N_1$, $M_1$, and $N_2$, $M_2$, respectively. Note that, $M_1$, $M_2$ need to be co-prime here while we have no restriction for $N_1$, $N_2$.
	The channel parameters satisfy the following conditions:
	\begin{equation}
		\label{eq30}
		\begin{aligned}
			l_{\tau_i} &\in [0,M_1M_2-1],\quad l_{\tau_i}\in \mathbb{Z} ,\\
			k_{\nu_i} &\in [-k_m,k_m], \quad k_{\nu_i}\in \mathbb{Z},
		\end{aligned}
	\end{equation}
	where the maximum Doppler shift tap $k_m < \min\{N_1,N_2\}/4$. 
	
	According to Eq. \eqref{eq9}, when only the delay is out of range, the receiving DD domain taps of $x[k_p,l_p]$ only spread along the Doppler axis with the range of $[k_p-k_m,k_p+k_m]$ but spread along the whole delay axis, respectively, without spreading over the whole DD domain.
	
\begin{figure}[H]
		\centering
		\includegraphics[width = 0.48\textwidth]{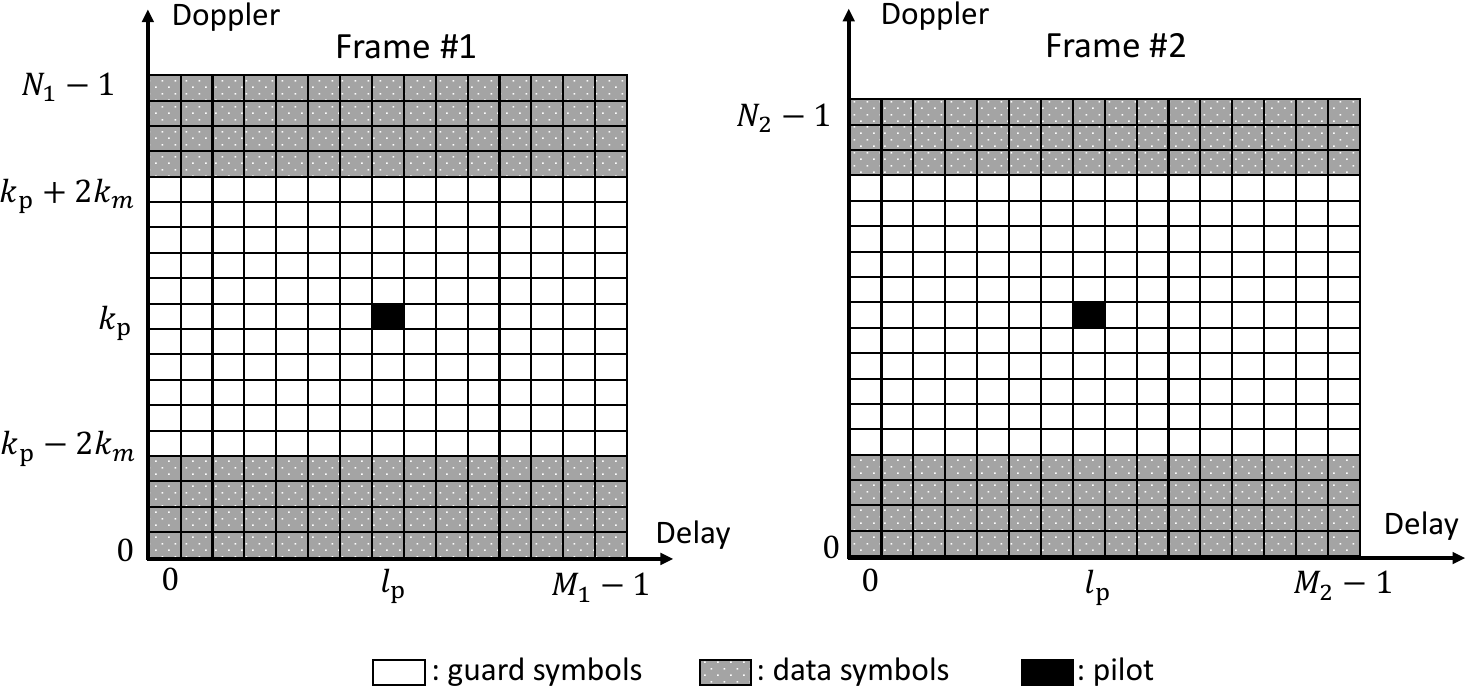}
		\caption{OTFS-ISAC frame with out-of-range delay.}
		\label{fig5}
	\end{figure}

	Therefore, to keep the communication symbols and sensing pilot from bothering each other, we arrange the pilot, data, and guard symbols in the DD domain as shown in Eq. \eqref{eq31} and Fig. \ref{fig5}:
	\begin{equation}
		\label{eq31}
		x[k,l]=\begin{cases}x_p&\quad k=k_p,l=l_p,\\0&\quad k_p-2k_m\leq k\leq k_p+2k_m,\\
			x_d[k,l]&\quad\text{otherwise,}\end{cases}
	\end{equation}
	
	where $x[k,l]$ is the transmit signal, $x_p$ is the pilot symbol, $x_d[k,l]$ is the data symbol, and $k_p$ and $l_p$ are the pilot indices. For simplicity, we consider the pilot location to be the same in each frame.
	
	Similarly, as analyzed in Sec. \ref{sec3.1}, the two frames can be separately processed to estimate the delay and Doppler taps of the scatterer as the threshold-based method in \cite{raviteja_embedded_2019}. 
	Their results can be noted as $\hat{l}_{\tau_1}$, $\hat{k}_{\nu_1}$ and $\hat{l}_{\tau_2}$, $\hat{k}_{\nu_2}$, respectively. 
	In this case, $\hat{k}_{\nu_1}$ and $\hat{k}_{\nu_2}$ will return the same true Doppler shifts, i.e., $\hat{k}_{\nu_1} = \hat{k}_{\nu_2}=\tilde{k}_{\nu_i}$. But the excessive delay will lead to periodic parameter estimation errors as shown in Eq. (\ref{eq14.2}). 
	
	One main advantage of this design is that the symbols for communication will not be affected because the out-of-range delay just introduces responses along the delay dimension, as analyzed in Eq. \eqref{eq4.6}.
	Furthermore, as we have discussed in Sec. \ref{sec3.3}, more frames can be used along the delay dimension. If $F$ subframes are used, then the unambiguous normalized delay range will be extended to $l_{\tau_i}\in[0,\mathrm{LCM}(M_1,M_2,...,M_F)-1]$.
	
	\subsection{Extend Doppler Range in OTFS-ISAC System with CRT}\label{sec3.4}
\begin{figure}
		\centering
		\includegraphics[width = 0.48\textwidth]{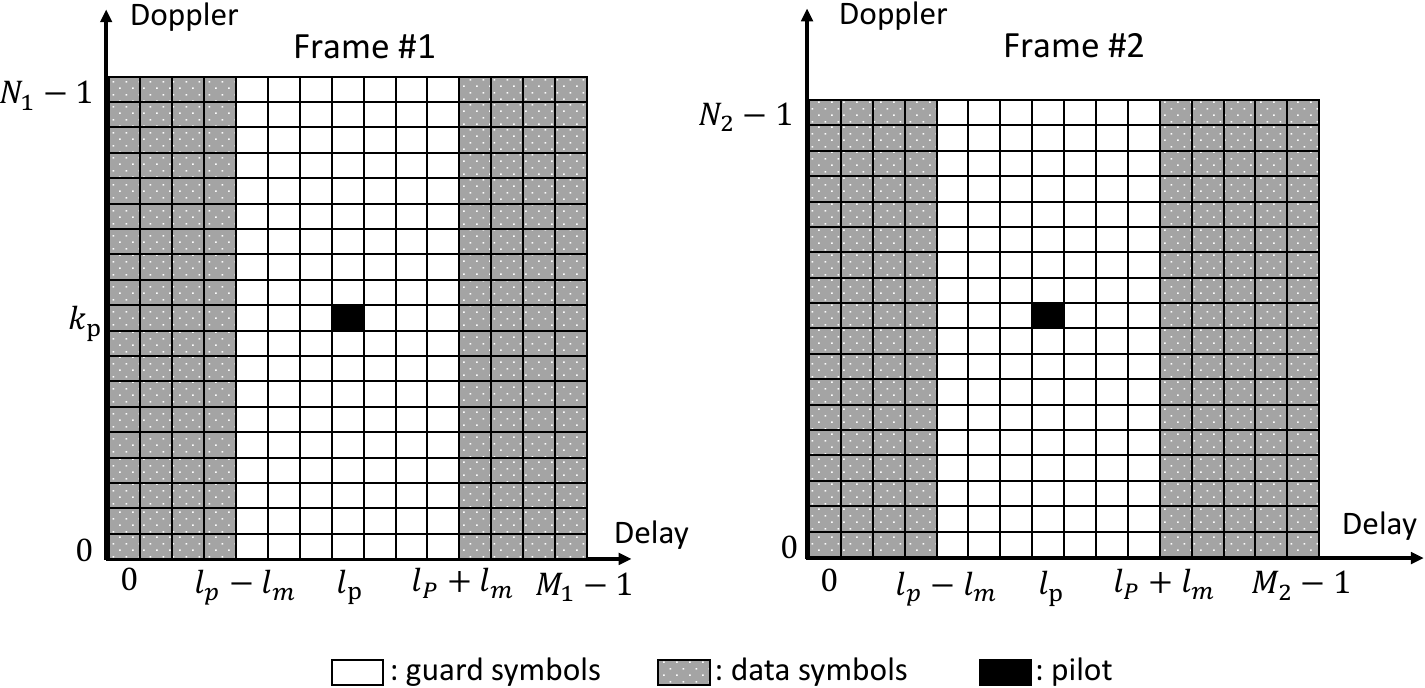}
		\caption{OTFS-ISAC frame with out-of-range Doppler.}
		\label{fig6}
	\end{figure}

	Similar to the previous section, if the target is with ultra-high speed but with in-range delay, i.e., only the Doppler shift is out of range, which may happen using a small subcarrier spacing to enhance spectral efficiency, the channel parameters will satisfy the following conditions:
	\begin{equation}
		\label{eq33}
		\begin{aligned}
			l_{\tau_i} &\in [0,l_m],\quad l_{\tau_i}\in \mathbb{Z} ,\\
			|k_{\nu_i}| &\in [0, (N_1N_2-1)/2], \quad k_{\nu_i}\in \mathbb{Z},
		\end{aligned}
	\end{equation}
	where $l_m < \min\{M_1,M_2\}/2$ is the maximum delay tap. 
	Likewise, the receiving DD domain taps of $x[k_p,l_p]$ only spread along the delay axis with the range of $[l_p-l_m,l_p+l_m]$ but spread along the whole Doppler axis. Hence the pilot, data, and guard symbols in the DD domain are arranged as shown in Eq. \eqref{eq34} and Fig. \ref{fig6}:
	\begin{equation}
		\label{eq34}
		\begin{aligned}
			x[k,l]=\begin{cases}x_p&\quad k=k_p,l=l_p,\\0&\quad l_p-l_m\leq l\leq l_p+l_m,\\
				x_d[k,l]&\quad\text{otherwise,}\end{cases}
		\end{aligned}
	\end{equation}
	
	As analyzed before, the excessive Doppler will lead to periodic parameter estimation error as shown in Eq. (\ref{eq14.1}). It will not affect the performance of communication because the communication symbols can be correctly demodulated from the received signals in Eq. \eqref{eq17} using the traditional method \cite{raviteja_embedded_2019}. Similarly, note their results as $\hat{l}_{\tau_1}$, $\hat{k}_{\nu_1}$ and $\hat{l}_{\tau_2}$, $\hat{k}_{\nu_2}$, respectively. 
	$\hat{l}_{\tau_1}=\hat{l}_{\tau_2}=\tilde{l}_{\tau_i}$ will return the same true delay shifts and the actual Doppler taps can be estimated using CRT as in Eq. (\ref{eq18a}). 
	When using more frames, the frame structure in Fig. \ref{fig6} will be extended to more subframes with co-prime time slot numbers like that in the last subsection.
	And the unambiguous normalized Doppler range are extended to $|k_{\nu_i}|\in [0,(\mathrm{LCM}(N_1,N_2,...,N_F))/2]$,where $F$ is the subframe number.
	
	\subsection{Extend Doppler and delay Range simultaneously}
	On the other hand, if we want to simultaneously extend the Doppler and delay detection range, the time slots and subcarriers in subframes should be co-prime and there will be no space for data symbols in the DD domain as the spread of the receiving taps. Thus the frame structure will be the same as discussed in Fig. \ref{fig2} for two and Fig. \ref{fig3} for more used subframes. In each frame, only one pilot is used and other symbols are set to zero. The actual parameters can be estimated using CRT as in Eq. (\ref{eq18a}) and Eq. (\ref{eq18b}) or Eq. (\ref{eq26}) and Eq. (\ref{eq27}).
	
	Note that in practice, the time and frequency bandwidth in two subframes may not be strictly equal when their numbers are simultaneously co-prime, which may lead to the fractional delay and Doppler taps. Although it will not affect the use of the proposed method, this phenomenon can be cleared off by using three subframes. Two of them have the same time duration and the other two have the same frequency bandwidth.
	
	\section{Simulation and Analysis}\label{sec5}
	Several simulations are performed to demonstrate the superiority of the proposed cross-frame parameter estimation framework over the traditional scheme. We compare the proposed framework with the classic whole-frame approach. 
	The classic whole-frame approach combines all frequency and time resources into one whole frame to enhance the sensing performance, as illustrated in Fig. \ref{fig3}. 
	We consider three Types of detections, i.e., Type 1: delay and Doppler both out of range, Type 2: delay out of range, and Type 3: Doppler out of range, respectively.
	
	The classic whole-frame approach has a finer delay and Doppler resolution, thus the representation of delay and Doppler in cross-frame approach are fractional. Therefore, fractional parameters are also considered in the simulations. Types of targets and cases of frame structures used in the simulations are summarized in Table \ref{tab2}.
	
	\begin{table}[ht]
		\caption{Types of detections and cases for simulation}
		\label{tab2}
		\centering
		\begin{tabular}{|c|p{0.6\linewidth}|}
			\hline
			\textbf{Condition} & \textbf{Description} \\
			\hline
			Type 1 &  both delay and Doppler out of range\\
			\hline
			Type 2 & only delay out of range\\
			\hline
			Type 3 & only Doppler out of range\\
			\hline
			Case 1 & Integer delay and Doppler taps in both whole frame and subframes \\
			\hline
			Case 2 &Integer delay and Doppler taps in whole frame, Integer delay and fractional Doppler in subframes\\
			\hline
			Case 3 & Fractional delay and Doppler taps in both whole frame and subframes\\
			\hline
		\end{tabular}
	\end{table}
	
\begin{figure}[htb]
		\centering
		\subfloat[Type 1]{\includegraphics[width = 0.3\textwidth]{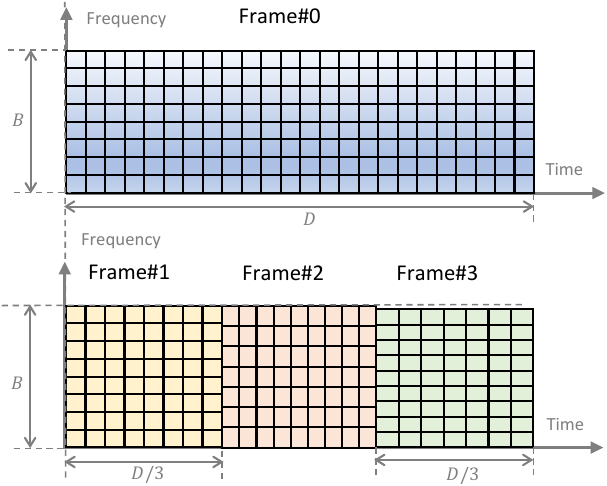}}\label{fig7a}
		\\
		\subfloat[Type 2]{\includegraphics[width = 0.22\textwidth]{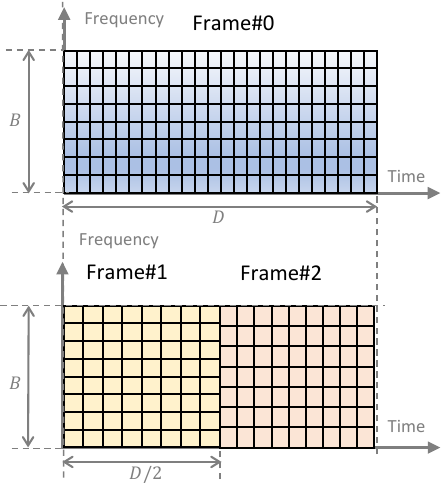}}
		\hfill
		\subfloat[Type 3]{\includegraphics[width = 0.22\textwidth]{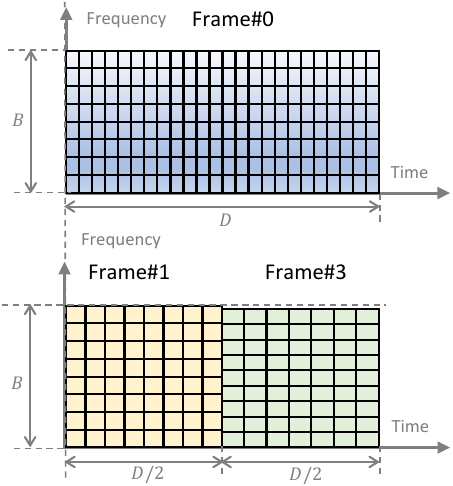}}
		\caption{Structure of Frames in Different Types.}
		\label{fig7}
	\end{figure}

	We use three subframes to estimate both excessive delay and Doppler (Type 1 in Table \ref{tab2}) to ensure the time and frequency bandwidths are strictly equal, where Frame 1 and Frame 2 occupy the same frequency width, Frame 1 and Frame 3 occupy the same time width, and the contrastive classic estimation method occupies all three frames, as shown in Fig. \ref{fig7}(a). 
	For the other two Types of targets, we use two subframes with same time or frequency bandwidth to estimate the excessive delay or Doppler, i.e., Frame 1 and Frame 2 are used for Type 2, shown in Fig. \ref{fig7}(b), and Frame 2 and Frame 3 for Type 3, shown in Fig. \ref{fig7}(c). 
	
	\subsection{Unambiguous delay and Doppler taps Comparison}
\begin{figure}
		\centering
		\includegraphics[width = 0.48\textwidth]{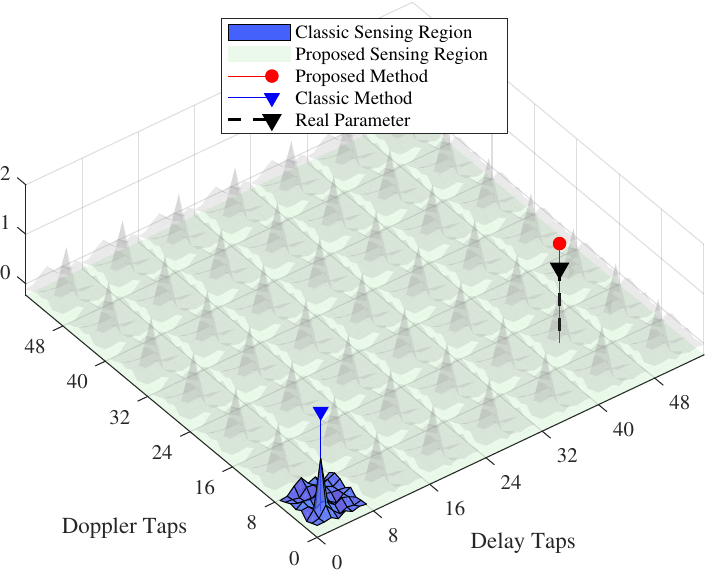}
		\caption{Unambiguous delay and Doppler taps of Two Methods.}
		\label{fig8}
	\end{figure}

	Fig. \ref{fig8} shows the max unambiguous delay and Doppler taps of the two frameworks at Type 1 and Case 1. 
	Frame 0 represents the whole frame in the classic method with $N_0 = 24$, $M_0 = 8$. Frame 1, 2, and 3 represent the three subframes to estimate the excessive channel parameters in the proposed method with $N_1 = 8, M_1 = 8, N_2 = 9, M_2 = 7$ and $ N_3 = 7, M_3 = 9$, respectively. The number of subcarriers and time slots are set small for display convenience.
	
	The blue surf in Fig. \ref{fig8} pictures the unambiguous delay and Doppler taps of the traditional method in \cite{raviteja_embedded_2019} with the sensing area of $M_1\times N_1 = 8\times 8 (\text{Delay Taps} \times \text{Doppler Taps})$ , while the green panel in the figure portrays the sensing area of the proposed method. 
	The sensing area becomes to $\mathrm{LCM}(M_1, M_2)\times \mathrm{LCM}(N_1, N_3)=56\times56(\text{Delay Taps} \times \text{Doppler Taps})$ in DD domain, 
	exhibiting the exponential gain in the unambiguous delay and Doppler taps, where $\mathrm{LCM}$ represents the least common multiple.
	
	The true delay and Doppler of the target are denoted by the black vertical bar, and the red and blue one represent the results of the proposed cross-frame framework and the classic whole-frame approach, respectively. 
	It can be found that the result of the classic whole-frame approach is limited in the blue surf where the response is periodically mapped in the area.
	While the proposed cross-frame framework can obtain the out-of-range parameters and the max unambiguous velocity and tolerable velocity is significantly enlarged. 
	
	\subsection{Sensing Performance in Statistic}
	Normalized Mean Square Error (NMSE) \cite{chani-cahuana_lower_2018} performance is used to analyze the accuracy of the proposed framework.
	\begin{equation}
		\label{eq35}
		\begin{aligned}
			\text{NMSE}(\theta) = \frac{\text{Var}(\hat{\theta}-\theta)}{\text{Var}(\theta)},
		\end{aligned}
	\end{equation}
	where $\theta$ denotes the real value of delay or Doppler taps, $\hat{\theta}$ is the estimated value, and $\text{Var}(\theta)$ is the variance of $\theta$. Note that, NMSE instead of MSE is used here, aiming to show the robustness of the framework for various delay and Doppler. Considering the estimated delays and Dopplers are normalized in the DD domain by 
	Eq. \eqref{eq10}, $\frac{1}{2}[\text{NMSE}(\text{delay})+\text{NMSE}(\text{Doppler})]$ is used as metrics to show the performance. 
	
	The practical system parameters follow that of \cite{raviteja_orthogonal_2019} and \cite{sturm_waveform_2011}, listed in Table \ref{tab3}.
	We consider a single target with unit channel gain and RCS, i.e., $h_1=1$, with its delay and Doppler being uniformly and randomly generated in ranges of Eq. (\ref{eq23}), (\ref{eq30}) and (\ref{eq33}), respectively. The Monte Carlo times are set to 100000.
	
	\begin{table}[tb]
		\caption{System Parameters}
		\label{tab3}
		\centering
		\begin{tabular}{|c|c|c|}
			\hline
			\textbf{Symbol} & \textbf{Parameter} & \textbf{Value} \\
			\hline
			\multicolumn{3}{|c|}{\textit{Common Settings}} \\
			\hline
			$f_c$ & Carrier frequency & 24GHz \\
			\hline
			$B$ & Bandwidth & 7.68 MHz \\
			\hline
			$\Delta f_0$ & Subcarrier spacing of whole frame& 30KHz\\
			\hline
			$M_0$ & Subcarrier number of whole frame & 256\\
			\hline
			$\Delta R_0$ & Range resolution of whole frame & 19.5m\\
			\hline
			$R_{\text{max}_0}$ &Max unambiguous range of whole frame& 5000m\\
			\hline
			$V_{\text{max}_0}$ &Max tolerable velocity of whole frame& $\pm$93.75m/s\\
			\hline
			\multicolumn{3}{|c|}{\textit{Type 1: delay and Doppler both out of range}} \\
			\hline
			$D_0$ & Duration of whole frame (frame 0)& 3.2ms\\
			\hline
			$N_0$ & Time slot number of whole frame & 96\\
			\hline
			$\Delta V_0$ & Velocity resolution of whole frame & 1.95m/s\\
			\hline
			$D_1$ & Duration of subframes (frame 1,2,3) & 1.07ms\\
			\hline
			$\Delta f_1$ & Subcarrier spacing of subframe 1& 30KHz\\
			\hline
			$M_1$ & Subcarrier number of subframe 1& 256\\
			\hline
			$N_1$ & Time slot number of subframe 1& 32\\
			\hline
			$\Delta f_2$ & Subcarrier spacing of subframe 2& 30.1177KHz\\
			\hline
			$M_2$ & Subcarrier number of subframe 2& 255\\
			\hline
			$N_2$ & Time slot number of subframe 2& 32\\
			\hline
			$\Delta f_3$ & Subcarrier spacing of subframe 3& 29.0625KHz\\
			\hline
			$M_3$ & Subcarrier number of subframe 3& 264\\
			\hline
			$N_3$ & Time slot number of subframe 3& 31\\
			\hline
			$\Delta R_s$ & Range resolution of subframes & 19.5m\\
			\hline
			$\Delta V_s$ & Velocity resolution of subframes& 5.86m/s\\
			\hline
			$R_{\text{max}_s}$ &Max unambiguous range (proposed method)& 127.5km\\
			\hline
			$V_{\text{max}_s}$ &Max tolerable velocity (proposed method) &$\pm$2.903km/s\\
			\hline
			\multicolumn{3}{|c|}{\textit{Type 2/3: only delay/Doppler out of range}} \\
			\hline
			$D_0$ & Duration of whole frame (frame 0)& 2.13ms\\
			\hline
			$N_0$ & Time slot number of whole frame & 64\\
			\hline
			$\Delta V_0$ & Velocity resolution of whole frame & 2.92m/s\\
			\hline
		\end{tabular}
	\end{table}

	\begin{figure}
		\centering
		\includegraphics[width = 0.4\textwidth]{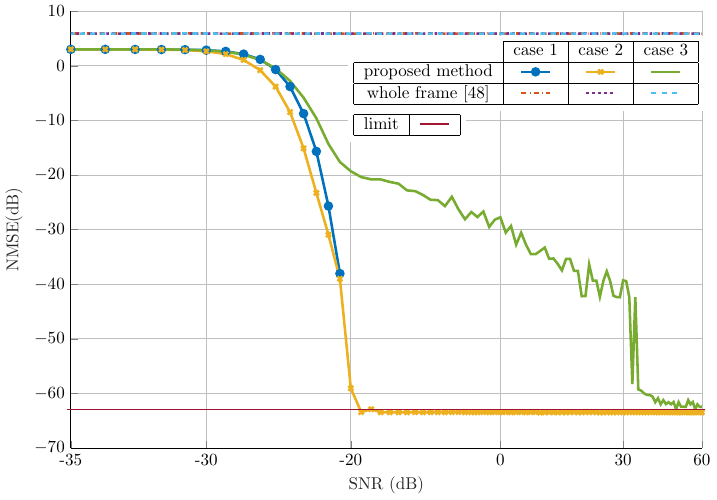}
		\caption{NMSE Performance in Type 1.}
		\label{fig9}
	\end{figure}
	
	\begin{figure}
		\centering
		\includegraphics[width = 0.4\textwidth]{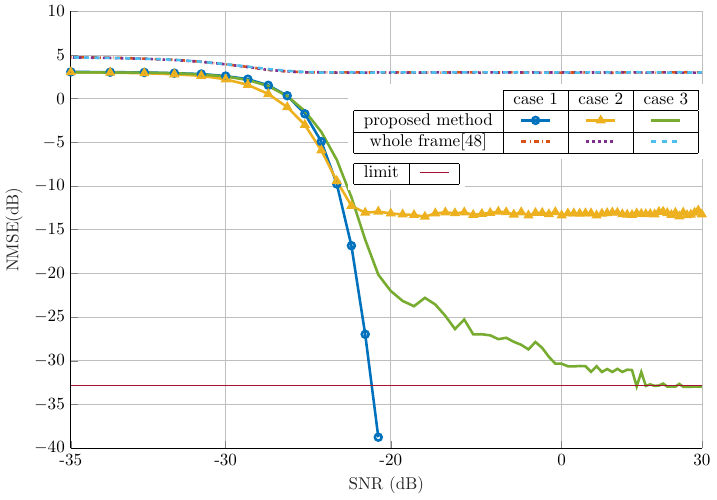}
		\caption{NMSE Performance in Type 2.}
		\label{fig10}
	\end{figure} 
	\begin{figure}
		\centering
		\includegraphics[width = 0.4\textwidth]{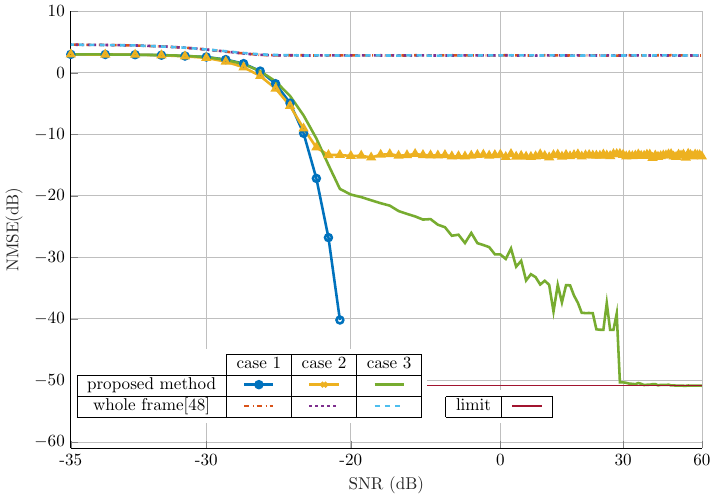}
		\caption{NMSE Performance in Type 3.}
		\label{fig11}
	\end{figure} 
	
	The three types with different cases in Table \ref{tab2} are considered in Fig. \ref{fig9}, \ref{fig10} and \ref{fig11}, respectively.
	Since the channel parameters are generated randomly, the NMSE results may fluctuate slightly due to the grid effect of the detection. The final estimation results are limited to integers, while the true delay and Doppler might be fractional. Thus a lower bound exists for case 2 and case 3. Usually, the bound for case 3 is lower and we show it in the figures with its legend denoted as 'limit'.
	
	It can be found in Fig. \ref{fig9}, \ref{fig10}, and \ref{fig11} that the classic method fails when the parameter is out of range.  
	The ambiguity in estimated results leads to an error whose value is a multiple of the period, independent from SNR. The proposed method can retrieve the multiple periods based on CRT, thus true delay and Doppler can be obtained.
	
	When there are no fractional delays and Dopplers, the NMSE of the cross-frame method will decrease to 0 as the SNR increases as shown in Case 1 of the three figures. The curves only display below -20 dB since the values have already reached 0 at a higher SNR.
	When fractional delays and Dopplers are considered, as shown in Cases 2 and 3 in these figures, the NMSE would degrade as SNR increases. The fluctuation can be attributed to the off-grid effects. When the SNR is large enough, the NMSE will decrease rapidly to a limit. This is because the closest integer to the true parameter can be achieved in overwhelming probability when SNR is sufficiently high.

	\section{Conclusion}\label{sec6}
	
	In this paper, a novel cross-frame OTFS parameter estimation method based on CRT is proposed to extend the max unambiguous range and tolerable velocity of the OTFS-ISAC system. 
	Specifically, the sensing ability can be extended by combining different OTFS subframes with co-prime subcarrier and time slot numbers, leveraging the periodicity of multicarrier waveforms in the delay-Doppler domain. 
	At the cost of a linear decrease in resolution, the proposed method can significantly extend the delay and Doppler range of the target detection on an exponential scale. 
	In practical applications, three kinds of frame structures are designed for ISAC scenarios including target at far distance, with ultra-high speed, or both of the situations.
	Simulation results show the effectiveness of the proposed method.
	However, one target scenario is considered in this paper. Multiple targets' scenarios are quite different and will be presented in our future works.

	\appendix[Derivation of Multicarrier Signal Periodicity]\label{app1}
	With one target considered, the sampled time-frequency domain received signal Eq. \eqref{EqSampling} can be written as:
	\begin{equation}
		\label{eq4.4}
		\begin{aligned}
			\hat{Y}[n,m]&=h_i\sum_{n^{\prime}=0}^{N-1}\sum_{m^{\prime}=0}^{M-1}\int X[n^{\prime},m^{\prime}]\\
			&g_{\mathrm{tx}}(t^{\prime}-\tau_i-n^{\prime}T)e^{j2\pi m^{\prime}\Delta f\left(t^{\prime}-\tau_i-n^{\prime}T\right)}\\
			& e^{j2\pi\nu_i\left(t^{\prime}-\tau_i\right)}g_{\mathrm{rx}}^{*}(t^{\prime}-nT)e^{-j2\pi m\Delta f\left(t^{\prime}-nT\right)}dt^{\prime}.
		\end{aligned}
	\end{equation}
	The out-of-range delay and Doppler can be expressed as an out-of-range part and an in-range part:
	\begin{equation}
		\label{eq4.3}
		\begin{aligned}
			&\tau_i=\alpha_i T+\hat{\tau}_{i},\quad\alpha_i \in \mathbb{Z},\quad\hat{\tau}_{i}<T,\\
			&\nu_i=\beta_i \Delta f+\hat{\nu}_{i},\quad\beta_i \in \mathbb{Z},\quad\hat{\nu}_{i}<\Delta f.
		\end{aligned}
	\end{equation}
	We consider that the pulse shaping filters at the transmitter and receiver occupy a time width of $T$, and a frequency width of $\Delta f$, respectively.
	To sample the received signal effectively, the received TF domain signal is written as:
	\begin{equation}
		\label{eqA2}
		\begin{aligned}
			Y[n,m] =& \hat{Y}[\alpha_i + n,\beta_i + m]\\
			=& h_{i}\sum_{n'=0}^{N-1}\sum_{m'=0}^{M-1}X[n^{\prime},m^{\prime}]e^{{j2\pi(\beta_{i}\Delta f+\widehat{\nu}_{i})n^{\prime}T}}\\
			&e^{{j2\pi\left((\beta_{i}\Delta f+\widehat{\nu}_{i})+m^{\prime}\Delta f\right)\left((n-n^{\prime})T-\widehat{\tau}_{i}\right)}}\\
			&A_{{g_{{\mathrm{rx}}}g_{{\mathrm{tx}}}}}((n-n^{\prime})T-\hat{\tau}_{i},(m-m^{\prime})\Delta f-\hat{\nu}_{i}),
		\end{aligned}
	\end{equation}
	where $n=0,1,...,N-1$, $m=0,1,...,M-1$.
	After SFFT, the DD domain received signal $y[k,l]$ can be expressed as:
	\begin{equation}
		\label{eq4.6}
		\begin{aligned}
			&y[k,l]=\frac{1}{\sqrt{NM}}h_{i}e^{{-j2\pi\hat{\tau}_{i}(\beta_{i}\Delta f+\hat{\nu}_{i})}}\sum_{n=0}^{N-1}\sum_{n'=0}^{M-1}\sum_{m=0}^{N-1}\sum_{m'=0}^{M-1}\\
			&e^{{j2\pi\hat{\tau}_{i}\cdot nT}}X[n^{\prime},m^{\prime}]e^{{j2\pi m^{\prime}\Delta f(nT-n^{\prime}T-\hat{\tau}_{i})}}\\
			&A_{{g_{{\mathrm{tx}}},g_{{\mathrm{rx}}}}}((n-n^{\prime})T-\hat{\tau}_{i},(m-m^{\prime})\Delta f-\hat{\nu}_{i})e^{{-j2\pi(\frac{nk}{N}-\frac{ml}{M})}},
		\end{aligned}
	\end{equation}
	where $k=0,1,...,N-1$, $l=0,1,...,M-1$. 
	% Note that due to the periodicity of DFT, the sampling rate is not affected by the excessive delay-Doppler in signal processing.
	
	\bibliographystyle{IEEEtran}
	\bibliography{reference.bib}
	
	\vfill
\end{document}